\documentclass[lettersize,journal]{IEEEtran}
\usepackage{amssymb}
\usepackage{makecell}
\usepackage{xcolor}
\usepackage{amsmath,amsfonts}
\usepackage{algorithmic}
\usepackage{algorithm}
\usepackage{array}
\usepackage[caption=false,font=normalsize,labelfont=sf,textfont=sf]{subfig}
\usepackage{textcomp}
\usepackage{stfloats}
\usepackage{url}
\usepackage{verbatim}
\usepackage{graphicx}
\usepackage{cite}
\usepackage{tikz}
\usetikzlibrary{arrows.meta, positioning, fit, backgrounds}

\begin{document}
\raggedbottom

\title{SecFD-FL: Frequency-Decoupled Federated Learning with Secure Aggregation for Charging Load Forecasting in Internet of Electric Vehicles}

\author{Mamunur Rashid,~\IEEEmembership{Member,~IEEE,} Nan Chen,~\IEEEmembership{Member,~IEEE,} Syed Ali Asad Rizvi,~\IEEEmembership{Member,~IEEE,} and Amiya Nayak,~\IEEEmembership{Fellow,~IEEE}
\thanks{Corresponding author: Nan Chen.}
\thanks{Mamunur Rashid and Syed Ali Asad Rizvi are with the Department of Electrical and Computer Engineering, Tennessee Technological University, Cookeville, TN, USA. (Email: mrashid44@tntech.edu, srizvi@tntech.edu)}
\thanks{Nan Chen and Amiya Nayak are with the School of Electrical Engineering and Computer Science, University of Ottawa, Ottawa, ON, Canada. (Email: nchen@uottawa.ca, nayak@uottawa.ca)}
}



\maketitle

\begin{abstract}
Accurate charging load forecasting is essential for the reliability and efficiency of the Internet of Electric Vehicles (IoEV), yet it is challenging on two fronts: charging demand is highly stochastic, and the data collected by distributed sensors is privacy-sensitive. Although federated learning (FL) preserves data privacy, collaborative training still remains vulnerable to an honest-but-curious server that can invert shared model updates to reconstruct private data.  
To address both data volatility and privacy concerns, we propose SecFD-FL, a frequency-decoupled federated learning framework for IoEV comprising: 1) a dual-stream gated recurrent unit (GRU) network that separately models short-term and long-term charging load volatility, and 2) a lightweight secure aggregation layer for privacy preservation. Specifically, the secure aggregation layer employs a zero-sum additive masking protocol that perturbs each client's upload with high-variance noise, preventing the server from reconstructing private data while preserving the exact aggregate.
Extensive experiments have been conducted on both EV-level and charging station-level using real datasets. The experiment results show that SecFD-FL reduces global RMSE by up to 60\% over centralized baselines, achieving Jain's fairness up to 0.991 across clients, and reduces DLG-attack reconstruction correlation by 8 times to 15 times relative to FedAvg — all with no additional per-round communication cost beyond a one-time key exchange.
\end{abstract}


\section{Introduction}
\IEEEPARstart{D}{riven} by the imperative climate change, the global transportation sector is undergoing a rapid transition toward electrification, with the worldwide stock of electric vehicles (EVs) projected to exceed 250 million units by 2030 \cite{ref_1}. However, range anxiety persists as a primary barrier to mass EV adoption, which necessitates strategic deployment of fast charging station (CS) networks \cite{ref_2}. 
Empowered by Internet of Things (IoT) technologies, EVs, CSs, and smart meters can interconnect to form the Internet of EVs (IoEV) to sense and collect EV data, which serves as a critical foundation for real-time operation at both edge and cloud ends \cite{new_ref_iotj2}. At the early EV adoption stage, high-granularity EV charging demand forecasting is critical. However, high-granularity forecasting is a challenging task as the charging demand profile can be volatile and spatially heterogeneous \cite{ref_3, ref_4}. Therefore, novel and effective forecasting approaches are urgently needed \cite{ref_5}.

In the literature, EV charging demand forecasting has evolved from statistical methods toward data-driven learning. Statistical models are inexpensive and interpretable but cannot capture the nonlinearity and stochasticity of EV loads \cite{ref_5_new}. Conventional machine learning (ML) methods, such as support vector machine and random forests (RF), improve nonlinear mapping but depend on hand-crafted features and struggle with long-range temporal dependencies in multi-scale charging behaviour. Deep learning (DL), a subclass that learns hierarchical representations directly from data, overcomes these limitations: recurrent architectures such as long short-term memory (LSTM) and gated recurrent unit (GRU) have become the dominant paradigm for extracting sequential patterns from charging demand \cite{ref_5}. However, deep models still rely on centralized data aggregation, which poses severe privacy risks to individual users, particularly in IoEV, where data are collected by geographically distributed IoT devices and transmitted over communication networks.

Federated learning (FL) emerges as a potential solution by enabling collaborative training that exchanges only local model updates rather than raw charging demand data \cite{ref_1}. FL has already been applied to EV-related forecasting tasks for privacy preservation, including blockchain-secured charging-load prediction \cite{ref_1}, decentralized charging-demand forecasting \cite{ref_20}, and clustered or transfer-learning-based station forecasting \cite{ref_21, ref_25}. However, existing works still have two critical limitations. 
First, feeding raw and highly volatile load signals directly into local models degrades forecasting accuracy, as these models struggle to capture rapid stochastic spikes of EV charging patterns \cite{ref_6}. Second, while FL avoids the direct transmission of raw data, the exchanged gradients remain vulnerable to inversion attacks, such as Deep Leakage from Gradients (DLG) \cite{ref_7}. Consequently, standard FL alone cannot simultaneously guarantee high forecasting accuracy and rigorous privacy protection.

To address aforementioned challenges, we propose a novel frequency-decoupled federated learning scheme for charging load forecasting in the IoEV. First, to achieve high-granularity EV charging forecasting, we introduce the Complete Ensemble Empirical Mode Decomposition with Adaptive Noise (CEEMDAN) method to decompose each participant's (an EV charging station or an individual EV) charging load into a set of intrinsic mode functions (IMFs) \cite{conference}. Through IMF analysis, we observe that EV charging loads can be divided into two streams: a high-frequency residual stream representing transient charging spikes and a low-frequency trend stream capturing seasonal and behavioral regularities. Therefore, a dual-stream  GRU network is designed to learn different frequency streams independently before combining them into a unified forecast. 
Second, to preserve client privacy during model exchange, we integrate a lightweight zero-sum additive masking protocol that delivers cryptographic-grade protection of local updates against gradient inversion attacks with no additional per-round communication overhead beyond a one-time key exchange. 
The principal contributions of this work are summarized as follows:
\begin{itemize}

\item We propose a novel dual-stream GRU network to capture both long-term and short-term EV charging demand trends leveraging CEEMDAN frequency decomposition to achieve a high-granularity forecasting.
\item We develop a lightweight zero-sum additive masking protocol for the data transmission during FL to safeguard local updates against gradient inversion, with its robustness empirically validated through a Deep Leakage from Gradients (DLG) attack.
\item We evaluate SecFD-FL on real-world station-level and user-level EV charging datasets against six representative federated baselines. SecFD-FL attains the highest accuracy (global $R^2$ of $0.927$ on EV~WATTS data and $0.847$ on Palo Alto data), while the secure-aggregation layer reduces DLG reconstruction correlation by $8.1$ times to $15.6$ times at no additional communication cost.
\end{itemize}

The remainder of this paper is organized as follows. Section II presents the related works. Section III introduces the system model and the proposed SecFD-FL framework. Section IV details the experimental setup, including the datasets, baseline models, and evaluation metrics. Section V presents the comparative results alongside a formal security analysis. Finally, Section VI concludes the paper with future work.

\section{Related Works}
Early efforts in charging demand forecasting for IoEV predominantly relied on classical ML algorithms applied to handcrafted features. Rathore et al. \cite{ref_8} benchmarked several regression-based learners, including RF, XGBoost, linear regression, artificial neural networks, and deep neural networks, for EV energy consumption prediction, with RF emerging as the most accurate model. Sun et al. \cite{ref_9} developed a support vector regression model trained on thirteen handcrafted features drawn from charging records, weather conditions, and calendar properties. More recently, Palaniappan et al. \cite{ref_10} introduced a stacked meta-fusion regressor that layers RF, XGBoost, and decision tree models to enhance predictive performance. Although these ML-based methods improved upon purely statistical baselines, they remain heavily reliant on manual feature engineering and exhibit limited capacity to capture both the long-range temporal dependencies and nonlinear dynamics inherent in EV charging behavior.
To overcome these limits, recent literature adopted DL architectures that automatically extract spatio-temporal features. Several studies couple convolutional and recurrent components for station-level forecasting, including Conv-LSTM/BiConv-LSTM encoder--decoders \cite{ref_11} and a CNN optimized by the niche-immunity lion algorithm \cite{ref_12}. Others incorporate graph structure: an edge-based graph attention network combined with LSTM \cite{ref_13}, and a graph-transformer that jointly models spatial and temporal patterns \cite{ref_14}. Further gains are reported from multi-feature fusion into LSTM inputs \cite{ref_16} and from a lite Transformer--GRU that fuses attention with sequential processing under dynamic conditions \cite{ref_17}. Despite their strong predictive performance, all of the aforementioned centralized DL frameworks require raw charging data to be aggregated at a central server. Such aggregation poses significant privacy and security risks, since EV charging records implicitly disclose users' mobility patterns, daily routines, and home or workplace locations. 

To address the privacy limitations of centralized training, FL has emerged as a promising paradigm in which clients collaboratively train a shared model without exposing raw data. Danish et al. \cite{ref_1} proposed Block-FedL, a blockchain-augmented FL framework that employs a Bidirectional LSTM for client-side load forecasting and replaces the centralized aggregator with on-chain smart contracts, thereby ensuring tamper-resistant and verifiable model updates. Wang et al. \cite{ref_20} further removed the central coordinator by introducing decentralized synchronous and asynchronous FL variants, the latter offering improved robustness in heterogeneous, real-world settings.
Several studies have integrated transfer learning into the FL pipeline to better cope with client heterogeneity. Wang et al. \cite{ref_21} proposed a domain-adaptive clustered federated transfer learning scheme that leverages maximum mean discrepancy to cluster clients with similar feature distributions and applies a multi-head attention transformer to capture salient spatio-temporal patterns. Other works have focused on architectural and security enhancements within the FL framework. Bian et al. \cite{ref_22} designed an attention-based convolutional model trained under FL, in which each service provider retains its time-series data locally and shares only model parameters. Wang et al. \cite{ref_23} proposed a lightweight authentication scheme for FL-driven energy prediction, although the contribution is primarily concerned with secure authentication rather than forecasting accuracy. Saputra et al. \cite{ref_24} developed an FL-based energy demand forecaster in which charging stations exchange model updates with a centralized provider server, while Perry et al. \cite{ref_25} introduced a clustering-driven FL approach that groups stations by behavioral similarity prior to aggregation.

Although existing FL-based methods successfully decouple model training from raw data sharing, several critical limitations remain unaddressed. First, existing frameworks feed raw load signals directly into the predictive backbone, neglecting the multi-scale, non-stationary nature of EV charging load in which transient and stochastic spikes coexist with long-term seasonal and behavioral trends. These components are difficult to learn jointly with a homogeneous architecture. Second, the shared gradients among all clients remain susceptible to gradient inversion attacks, such as DLG, which can reconstruct private samples from observed updates. However, this vulnerability is rarely examined or mitigated in prior studies. Third, comprehensive benchmarking against state-of-the-art federated optimizers (e.g., FedProx, FedAdam, SCAFFOLD, FedDyn, Per-FedAvg) on heterogeneous, CS-level and EV-level datasets is largely absent, leaving the relative merits of competing FL designs in the EV forecasting context unclear. The proposed SecFD-FL framework directly addresses these gaps by 1) developing a dual-stream GRU to capture both short-term and long-term load trends; 2) embedding a lightweight zero-sum additive masking protocol to defend DLG attacks; 3) conducting extensive multi-level benchmarking against representative federated baselines.

\section{Proposed SecFD-FL Framework}
\label{sec:methodology}

\subsection{System Model and Problem Formulation}
\label{sec:sysmodel}

\textbf{System model.} We consider a federation of $K$ clients coordinated by a single central server that collaboratively train a shared charging-load forecasting model without exchanging raw data. The framework can be generalized and evaluated at different deployment levels. At the \emph{station level}, each client is an EV charging station and the server is operated by the charging-network operator; the trained model supports station-side load monitoring and operation planning. At the \emph{user level}, each client is an individual EV (or its associated user account) and the server is operated by the charging-service or fleet operator; here the model supports user- and fleet-level demand anticipation. In both settings, every client trains locally on its own data and uploads only model updates; the server's role is restricted to aggregating these updates and broadcasting the result, and it never accesses raw charging records.

\textbf{Problem formulation.} Each client $k\in\{1,\dots,K\}$ holds a private dataset $\mathcal{D}_k=\{(\mathbf{x}_{k,i},\mathbf{y}_{k,i})\}_{i=1}^{n_k}$ 
of $n_k$ samples, where $\mathbf{x}_{k,i}\in\mathbb{R}^{\tau\times M}$ is a $\tau$-step lookback window over $M$ frequency channels and $\mathbf{y}_{k,i}\in\mathbb{R}^{M}$ is the next-step target. Let $n=\sum_{k=1}^{K} n_k$ denote the total number of training samples across the federation. The goal is to learn a single global model $f_\mathbf{w}$ that minimizes the average next-step charging-load forecasting error, measured as the mean squared error (MSE) between predictions and ground truth. Denoting client $k$'s local forecasting loss as
\begin{equation}
\mathcal{L}_k(\mathbf{w})=\frac{1}{n_k}\sum_{i=1}^{n_k}\bigl\|f_\mathbf{w}(\mathbf{x}_{k,i})-\mathbf{y}_{k,i}\bigr\|_2^{2},
\label{eq:localloss}
\end{equation}
the federation solves the data-weighted global objective
\begin{equation}
\min_{\mathbf{w}}\;\mathcal{L}(\mathbf{w})=\sum_{k=1}^{K}\frac{n_k}{n}\,\mathcal{L}_k(\mathbf{w}),
\label{eq:fedobj}
\end{equation}
without sharing any $\mathcal{D}_k$ with the server. Here, $\mathcal{L}_k$ is the local average MSE over client $k$'s $n_k$ samples, and the weight $n_k/n$ scales each client's contribution by its share of the total data, so that clients holding more data influence the global model proportionally. Equivalently, \eqref{eq:fedobj} is the empirical risk minimized over the union of all clients' data.

\textbf{Threat model.} We adopt an \emph{honest-but-curious} central server (the aggregating entity of Section~\ref{sec:sysmodel}, e.g., the charging-network or service operator): it executes the aggregation protocol faithfully but may attempt to infer clients' private data from the updates it receives. The
representative attack we consider is DLG \cite{ref_7}, in which the server
inverts an observed update to reconstruct the underlying
training samples.

\subsection{System Architecture of the SecFD-FL Framework}
\begin{figure*}[t]
\centering
\includegraphics[width=7 in]{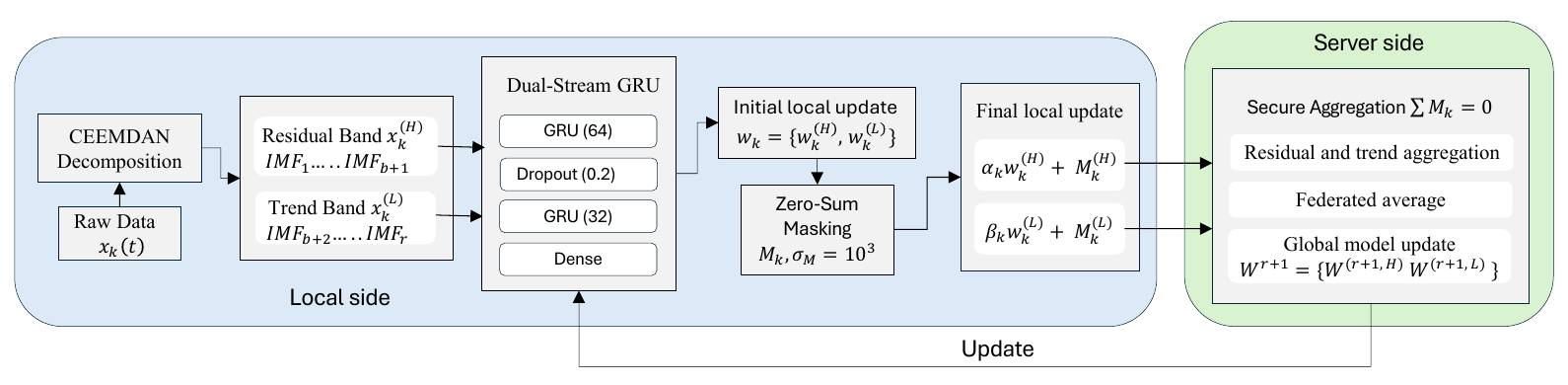}
\caption{System Architecture of the SecFD-FL framework.}
\label{full_method}
\end{figure*}

The proposed SecFD-FL framework is divided into localized client-side processing and centralized server-side orchestration.
As illustrated in Fig.~\ref{full_method}, the process begins locally by decomposing raw load data $x_{k}(t)$ via CEEMDAN into a high-frequency residual band $\mathbf{x}_k^{(H)}$ and a lower-frequency trend band $\mathbf{x}_k^{(L)}$. These decoupled bands are processed through a dual-stream GRU network to extract features and generate an initial local model update $w_{k}$. To protect client privacy prior to transmission, a zero-sum masking mechanism is applied to form a secure final local update. On the server side, these masked updates are securely aggregated; because the masks sum to zero ($\sum M_{k}=0$), the server can perform a standard federated average for both residual and trend components without exposing raw local data. Finally, the newly aggregated global model $W^{r+1}$ is broadcast back to the clients for the next training round.

\subsection{CEEMDAN-Based Frequency Decoupling}
\label{sec:ceemdan}

EV charging load exhibits slow seasonal trends with sharp transient bursts. Feeding such non-stationary data to a single learner mixes incompatible scales and degrades accuracy. In this paper, we leverage the Empirical Mode Decomposition (EMD) to decompose a nonlinear signal into a set of simpler oscillatory components called IMFs. CEEMDAN refines EMD by repeatedly adding low-amplitude noise and averaging the resulting decompositions, which reduces mode mixing and yields more stable, well-separated IMFs \cite{ref_26}. Each client therefore decomposes its raw load $x_k(t)$ via CEEMDAN, yielding $M_k$ IMFs $\tilde{c}_{k,m}(t)$ and a residual signal $r_{k,M}(t)$:

\begin{equation}
x_k(t)=\sum_{m=1}^{M_k}\tilde{c}_{k,m}(t)+r_{k,M_k}(t),
\label{eq:ceemdan_recon}
\end{equation}

\begin{equation}
\tilde{c}_{k,m}(t)=\tfrac{1}{I}\sum_{i=1}^{I}
E_m\!\bigl[r_{k,m-1}(t)+\varepsilon_m
E_{m-1}(\varepsilon^{(i)}(t))\bigr],
\label{eq:ceemdan_imf}
\end{equation}
where $E_m[\cdot]$ extracts the $m$-th mode, $\varepsilon^{(i)}\!\sim\!\mathcal{N}(0,1)$ is white noise, and $\varepsilon_m$ controls noise amplitude. IMFs are ordered by decreasing instantaneous frequency. 
Since $M_k$ varies across clients, each decomposition is zero-padded to a common dimensionality $M$, 
so that every client's input has a uniform shape. The frequency boundary $b$ then partitions this fixed-length, padded representation:
A frequency boundary $b$ partitions the channels into a
high-frequency \emph{residual} band and a low-frequency
\emph{trend} band:
\begin{align}
\mathbf{x}_k^{(H)}(t)&=\bigl[\tilde{c}_{k,1}(t),\dots,\tilde{c}_{k,b+1}(t)\bigr],\label{eq:hbn}\\
\mathbf{x}_k^{(L)}(t)&=\bigl[\tilde{c}_{k,b+2}(t),\dots,\tilde{c}_{k,M}(t),r_{k,M}(t)\bigr].\label{eq:lbn}
\end{align}
Here, $\mathbf{x}_k^{(H)}$ captures high-frequency transient dynamics (charging spikes, irregular usage) and $\mathbf{x}_k^{(L)}$ captures low-frequency trend patterns (daily/weekly cycles). The original signal $x_k(t)$ can be reconstructed by summing both high-frequency and low-frequency IMFs.
Each IMF channel is normalized with a min-max scaler fitted on training data:
\begin{equation}
\hat{c}_{k,m}(t)=\frac{\tilde{c}_{k,m}(t)-\min(\tilde{c}_{k,m}^{\text{tr}})}
{\max(\tilde{c}_{k,m}^{\text{tr}})-\min(\tilde{c}_{k,m}^{\text{tr}})},
\label{eq:scale}
\end{equation}
Supervised training sequences are constructed using a sliding window of length $\tau$,  giving inputs $\mathbf{x}_{k,i}\!\in\!\mathbb{R}^{\tau\times M}$
and targets $\mathbf{y}_{k,i}\!\in\!\mathbb{R}^{M}$:
\begin{align}
\mathbf{x}_{k,i} &= [\hat{\mathbf{c}}_k(t-\tau+1), \ldots, \hat{\mathbf{c}}_k(t)] \in \mathbb{R}^{\tau \times M}, \label{eq:slide_window_1}\\
\mathbf{y}_{k,i} &= \hat{\mathbf{c}}_k(t+1) \in \mathbb{R}^{M}.
\label{eq:slide_window_2}
\end{align}

\subsection{Dual-Stream GRU Networks}
\label{sec:dualgru}
\begin{figure}[t]
\centering
\includegraphics[width=2.5 in]{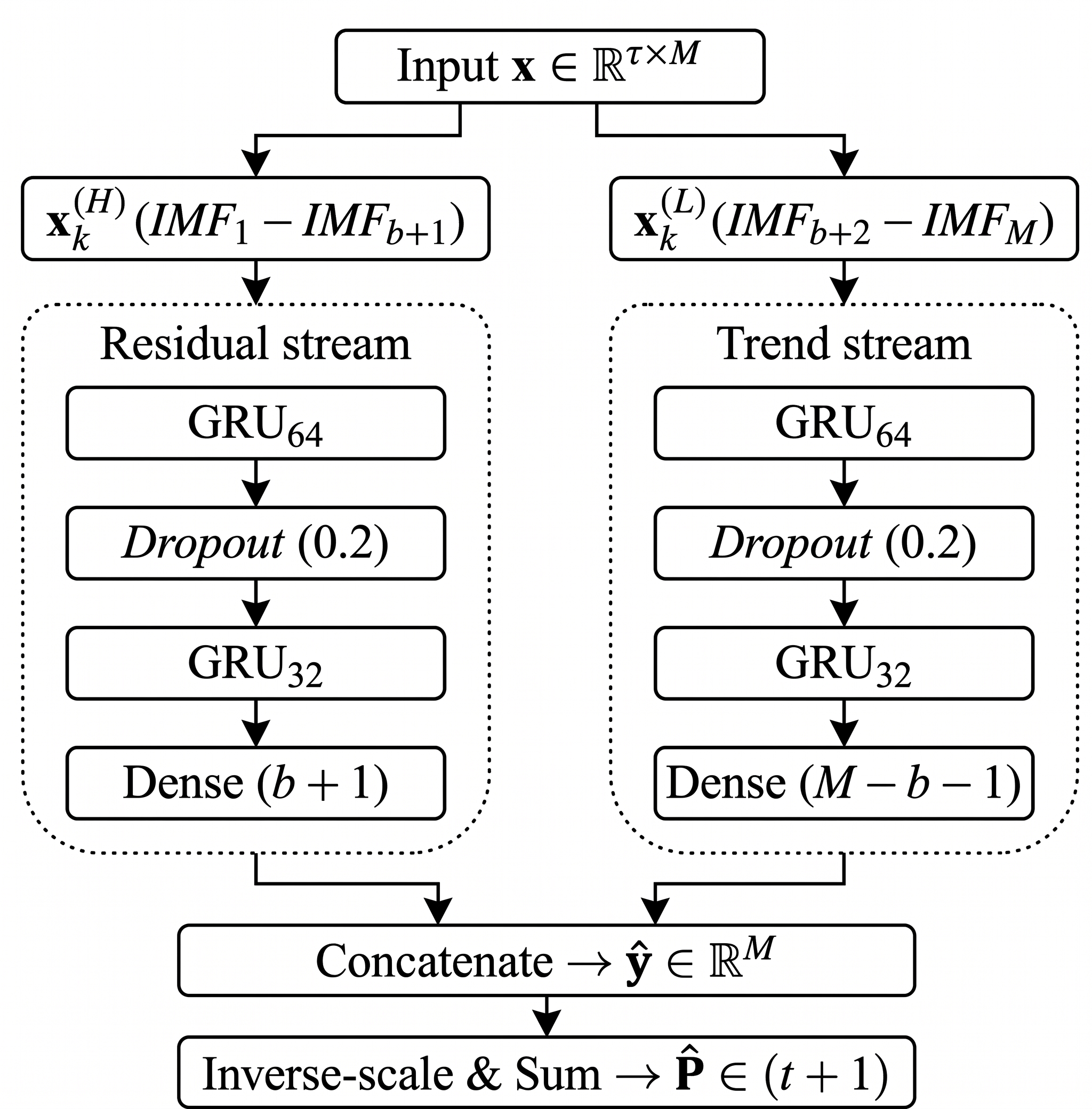}
\caption{Dual-stream GRU. A parameter-free Lambda layer routes the high-frequency residual band ($\mathbf{x}_k^{(H)}$) and low-frequency trend band ($\mathbf{x}_k^{(L)}$) into two parameter-disjoint GRU branches, whose outputs are concatenated and inverse-scaled to reconstruct the next-step load $\hat{P}(t{+}1)$.}
\label{fig:dualgru}
\end{figure}

The framework uses a lambda layer to split the multi-channel input $\mathbf{x}$ into two parallel streams. This parameter-free slicing ensures that each network branch handles a specific frequency sub-band independently, as illustrated in Fig.~\ref{fig:dualgru}.
The input is routed as follows: 
\begin{equation}
\mathbf{x}_k^{(H)}\!=\mathbf{x}[:,:,0{:}b{+}1],\quad
\mathbf{x}_k^{(L)}\!=\mathbf{x}[:,:,b{+}1{:}M].
\label{eq:slice}
\end{equation}

Each branch $s\!\in\!\{H,L\}$ uses a stacked GRU \cite{ref_30} with intermediate dropout:
\begin{align}
\mathbf{h}_t^{(s,1)}&=\mathrm{GRU}_{64}^{(s,1)}\!\bigl(\mathbf{x}_t^{(s)},\mathbf{h}_{t-1}^{(s,1)}\bigr),\label{eq:gru1}\\
\mathbf{h}_\tau^{(s,2)}&=\mathrm{GRU}_{32}^{(s,2)}\!\bigl(\mathrm{Drop}_{0.2}(\mathbf{h}_t^{(s,1)}),\mathbf{h}_{t-1}^{(s,2)}\bigr),\label{eq:gru2}\\
\hat{\mathbf{y}}^{(s)}&=\mathbf{W}_o^{(s)}\mathbf{h}_\tau^{(s,2)}+\mathbf{b}_o^{(s)},\label{eq:dense}
\end{align}
The forecasted IMF vector is given by: $\hat{\mathbf{y}}=[\hat{\mathbf{y}}^{(H)},\hat{\mathbf{y}}^{(L)}]\!\in\!\mathbb{R}^{M}$.
The final load is reconstructed by inverse-scaling and summing all predicted IMF channels:
\begin{equation}
\hat{P}(t{+}1)=\sum_{m=1}^{M}\mathrm{InverseScale}_m\!\bigl(\hat{y}_m\bigr).
\label{eq:invsum}
\end{equation}
Since the two branches share no weights, the global parameter vector $\mathbf{w}$ is partitioned into disjoint subsets:  
\begin{equation}
\mathbf{w}=\{\mathbf{w}^{(H)},\mathbf{w}^{(L)}\},\quad
\mathbf{w}^{(H)}\cap\mathbf{w}^{(L)}=\varnothing,
\label{eq:disjoint}
\end{equation}
This architectural independence enables heterogeneous aggregation to independently update the residual and trend components.
\subsection{Heterogeneous Federated Aggregation}
\label{sec:hetagg}
Since the two streams capture statistically distinct
phenomena, applying identical aggregation rules is suboptimal.
We therefore design a stream-specific aggregation policy.
\subsubsection{Stream-specific aggregation weights}
For the residual stream, \emph{Data-proportional} weights $\alpha_k$ are applied, as stations with more data samples produce statistically stronger transient estimates:
\begin{equation}
\alpha_k=\frac{n_k}{\sum_{j=1}^{K}n_j}.
\label{eq:alpha}
\end{equation}
\emph{Spectral-similarity} weights govern the trend stream.
During the one-time initialization phase, 
each client computes the trend-band spectral signature of its mean training feature $\bar{\mathbf{f}}_k\!\in\!\mathbb{R}^{M-b-1}$ via Fast Fourier Transform (FFT) magnitude normalization. The FFT is used only to derive these aggregation-weighting signatures; it is independent of CEEMDAN, which performs the time-domain signal decomposition through empirical mode sifting as defined below: 
\begin{equation}
\mathbf{s}_k=\frac{|\mathrm{FFT}(\bar{\mathbf{f}}_k)|}{\bigl\||\mathrm{FFT}(\bar{\mathbf{f}}_k)|\bigr\|_2}.
\label{eq:sig}
\end{equation}
By sharing normalized signatures with the server, the server forms the pairwise cosine-similarity graph
$S_{kj}=\max(0,\mathbf{s}_k^{\!\top}\mathbf{s}_j/(\|\mathbf{s}_k\|\|\mathbf{s}_j\|+\epsilon))$. The final trend weight is then derived from each client's mean similarity centrality:
\begin{equation}
\beta_k=\frac{\bar{S}_k}{\sum_{j}\bar{S}_j},\quad
\bar{S}_k=\tfrac{1}{K}\sum_{j=1}^{K}S_{kj}.
\label{eq:beta}
\end{equation}

Importantly, sharing  $\mathbf{s}_k$ preserves strict data privacy as it reveals no time-domain or sample-level information. $\beta_k$ is the trend-stream aggregation weight, computed entirely server-side from the similarity graph $S_{kj}$. 

\subsubsection{Stream-specific aggregation rules}
At communication round $r$, after each client trains $E$ local epochs, the
server applies stream-specific aggregation:
\begin{align}
\mathbf{w}^{(L,r)}&=\sum_{k=1}^{K}\beta_k\,\mathbf{w}_k^{(L,r)},\label{eq:aggL}\\
\mathbf{w}^{(H,r)}&=(1{-}\gamma)\sum_{k=1}^{K}\alpha_k\,\mathbf{w}_k^{(H,r)}+\gamma\,\mathbf{w}^{(H,r-1)},\label{eq:aggH}
\end{align}
where $\gamma\!\in\![0,1]$ is the local-retention ratio.
The momentum-like retention term in Eq.~\eqref{eq:aggH} preserves a
fraction of the previous global residual, preventing
station-specific transient patterns from being washed out by aggressive global averaging, a known weakness of standard FedAvg under highly heterogeneous and non-IID conditions \cite{ref_31}.

\subsection{Zero-Sum Secure Aggregation and Privacy Analysis}
\label{sec:secagg}
Equations~\eqref{eq:aggL} - \eqref{eq:aggH} require the server
to read each $\mathbf{w}_k^{(s)}$, exposing them to gradient
inversion. To prevent privacy leakage, we embed a lightweight additive masking protocol that lets the server compute the exact aggregate
without observing individual updates.

\subsubsection{Mask construction}
Each pair of clients $(i,j)$, $i\neq j$, shares a secret seed $s_{ij}=s_{ji}$ via a one-time pairwise key exchange. 
Client $k$ generates its mask $\mathbf{M}_k$ entirely locally as the signed sum of pairwise pseudorandom vectors $\mathrm{PRG}(s_{kj})$:
\begin{equation}
\mathbf{M}_k=\sum_{j\neq k}\mathrm{sgn}(k-j)\,\mathrm{PRG}(s_{kj})
\label{eq:mask1}
\end{equation}
\begin{equation}
\mathrm{PRG}(s_{kj})\sim\mathcal{N}\!\left(\mathbf{0},\tfrac{\sigma_M^2}{K-1}\mathbf{I}\right),
\label{eq:mask2}
\end{equation}
where $\mathrm{sgn}(k-j)=+1$ if $k<j$ and $-1$ otherwise. Since $s_{kj}=s_{jk}$, clients $k$ and $j$ derive the identical vector $\mathrm{PRG}(s_{kj})$ but apply it with opposite sign, so every pairwise term cancels between exactly two clients and the masks sum to zero across the federation:
\begin{equation}
\sum_{k=1}^{K}\mathbf{M}_k=\mathbf{0}.
\label{eq:zerosum}
\end{equation}
Each client's mask has element-wise variance $\sigma_M^2$ (the $K{-}1$ pairwise terms each contribute $\sigma_M^2/(K{-}1)$), with $\sigma_M{=}10^3$, to match a high magnitude above the natural weight scale, which is consistent with the SNR in Eq.~\eqref{eq:snr}. Because $\mathbf{M}_k$ is computed solely from seeds $k$ shares pairwise with the other $K{-}1$ clients, no client can reconstruct another client's mask, and the server observes only the masked contributions of Eq.~\eqref{eq:upload}, never an individual $\mathbf{M}_k$.

\subsubsection{Masked upload and recovery}
Each client computes its pre-weighted model update and adds its mask before uploading to the server. For stream $s$ with aggregation weight $\omega_k^{(s)}$, client $k$ uploads:
\begin{equation}
\hat{\mathbf{c}}_k^{(s)}=\omega_k^{(s)}\mathbf{w}_k^{(s)}+\mathbf{M}_k^{(s)}.
\label{eq:upload}
\end{equation}
The server then sums the masked contributions as:
\begin{equation}
\sum_{k=1}^{K}\hat{\mathbf{c}}_k^{(s)}
=\sum_{k=1}^{K}\omega_k^{(s)}\mathbf{w}_k^{(s)}
+\!\!\!\underbrace{\sum_{k=1}^{K}\mathbf{M}_k^{(s)}}_{=\,\mathbf{0}\;\text{by~\eqref{eq:zerosum}}}\!\!\!.
\label{eq:recover}
\end{equation}
Since the masks cancel exactly, the convergence trajectory of SecFD-FL is bit-identical to its unmasked counterpart, ensuring robust privacy at zero utility cost.

\subsubsection{Privacy guarantee against DLG}
In the standard federated learning, an adversary can recover private data through DLG. 
This attack reconstructs the original input $\mathbf{x}$ by iteratively minimizing the distance between the observed gradient $\mathbf{g}_{\text{obs}}$ and a synthetic gradient generated from dummy data $(\mathbf{x}', \mathbf{y}')$, where $\mathcal{L}$ is the same mean squared forecasting loss minimized during training (Eq.~\eqref{eq:fedobj}):
\begin{equation}
\mathbf{x}^{*}=\arg\!\min_{\mathbf{x}'} 
\bigl\|\nabla_\mathbf{w}\mathcal{L}(\mathbf{w};\mathbf{x}',\mathbf{y}')-\mathbf{g}_\text{obs}\bigr\|_2^2,
\label{eq:dlg}
\end{equation}
Under the proposed zero-sum secure aggregation defense, the server no longer observes raw gradients, but rather the masked contributions
$\hat{\mathbf{c}}_k^{(s)}$, whose signal-to-noise ratio is
\begin{equation}
\mathrm{SNR}_k\!=\!\tfrac{\|\omega_k^{(s)}\mathbf{w}_k^{(s)}\|_2}{\|\mathbf{M}_k^{(s)}\|_2}
\!\approx\!\tfrac{\omega_k\sigma_w}{\sigma_M}.
\label{eq:snr}
\end{equation}
Using our system parameters ($\omega_k\!\approx\!0.2,\sigma_w\!\approx\!10^{-2},\sigma_M\!=\!10^{3}$), the resulting
$\mathrm{SNR}_k\!\approx\!2{\times}10^{-6}$. This indicates that the true gradient signal is buried five orders of magnitude below the pseudorandom noise. Consequently, any DLG optimization in Eq.~\eqref{eq:dlg} performed on the masked update will converge to the random mask rather than the private data features, making the reconstruction statistically uncorrelated with the original charging profiles. 
\subsection{Overall Algorithm}
\label{sec:algorithm}
Algorithm~\ref{alg:fedcfd} formalizes the end-to-end orchestration of the proposed SecFD-FL framework, integrating one-time signal decomposition with an iterative, secure federated training loop. The procedure employs a dual-stream architecture to independently update stochastic residuals and seasonal trends, utilizing stream-specific weights to mitigate client drift while preserving privacy through zero-sum masking.

\begin{algorithm}[]
\caption{{SecFD-FL: Frequency-Decoupled FL with Secure Aggregation for IoEV}}\label{alg:fedcfd}
\begin{algorithmic}[1]
\STATE \textbf{Input:} Clients $\{\mathcal{D}_k\}_{k=1}^{K}$; rounds $T$; local epochs $E$; batch size $B$; learning rate $\eta$; boundary $b$; retention $\gamma$; mask scale $\sigma_M$.
\STATE \textbf{Output:} Global model $\mathbf{w}^{(T)}$.
\STATE 
\FOR{$k=1,\dots,K$}
    \STATE Apply CEEMDAN to obtain $M$ IMFs (Eq.~\eqref{eq:ceemdan_imf})
    \STATE Partition by $b$ into $\mathbf{x}_k^{(H)},\mathbf{x}_k^{(L)}$ (Eqs.~\eqref{eq:hbn}--\eqref{eq:lbn})
    \STATE Min-Max scale each IMF channel (Eq.~\eqref{eq:scale}); build $\tau$-step sequences
    \STATE Compute spectral signature $\mathbf{s}_k$ (Eq.~\eqref{eq:sig}) and send to server
\ENDFOR
\STATE Compute $\{\alpha_k\},\{\beta_k\}$ (Eqs.~\eqref{eq:alpha},\eqref{eq:beta}); init $\mathbf{w}^{(0)}$; key-exchange for mask seeds
\STATE 
\FOR{$r=1,\dots,T$}
    \STATE Server broadcasts $\mathbf{w}^{(r-1)}$
    \FOR{\textbf{in parallel} $k=1,\dots,K$}
        \STATE $\mathbf{w}_k\!\leftarrow\!\mathbf{w}^{(r-1)}$; train $E$ epochs on $\mathcal{D}_k$
        \STATE Generate zero-sum masks $\mathbf{M}_k^{(H)},\mathbf{M}_k^{(L)}$ using Eq.~\eqref{eq:mask1}
        \STATE Upload $\hat{\mathbf{c}}_k^{(s)}\!=\!\omega_k^{(s)}\mathbf{w}_k^{(s)}+\mathbf{M}_k^{(s)}$ using Eq.~\eqref{eq:upload}
    \ENDFOR
    \STATE $\mathbf{w}^{(L,r)}\!\leftarrow\!\sum_{k}\hat{\mathbf{c}}_k^{(L)}$ 
    \STATE $\tilde{\mathbf{w}}^{(H,r)}\!\leftarrow\!\sum_{k}\hat{\mathbf{c}}_k^{(H)}$
    \STATE $\mathbf{w}^{(H,r)}\!\leftarrow\!(1{-}\gamma)\tilde{\mathbf{w}}^{(H,r)}+\gamma\,\mathbf{w}^{(H,r-1)}$ 
    \STATE $\mathbf{w}^{(r)}\!\leftarrow\!\{\mathbf{w}^{(H,r)},\mathbf{w}^{(L,r)}\}$
\ENDFOR
\STATE 
\STATE Predict $\hat{\mathbf{y}}\!=\![\hat{\mathbf{y}}^{(H)},\hat{\mathbf{y}}^{(L)}]$; recover $\hat{P}(t{+}1)$ using Eq.~\eqref{eq:invsum}.
\end{algorithmic}
\end{algorithm}


\section{Experimental Settings}

\subsection{Datasets}
\label{sec:dataset}

To assess SecFD-FL across both granularities at which charging
forecasting is typically deployed in IoEV, we use one \emph{station-level} dataset from City of Palo Alto and one \emph{user-level} dataset from EV WATTS. 

\subsubsection{Station-level data — City of Palo Alto}
The first dataset is the publicly released EV charging records of Palo Alto \cite{ref_27}, which logs every charging session at municipally operated stations, including
session start time, charging duration, and energy delivered (kWh). The dataset comprises five municipally operated charging stations, identified by their media access control (MAC) addresses; we use all five as federated clients ($K=5$), each containing multi-year session histories.

\subsubsection{User-level data — EV WATTS}
The second dataset is the EV WATTS \cite{ref_28}, which logs per-vehicle charging events
across a heterogeneous national fleet. To ensure each client has sufficient history for sequence modelling, we retain only vehicles with at least 300 recorded sessions \emph{and} a lifespan of at
least 365 days, yielding 35 eligible EVs. For a fair comparison with the station-level dataset, we randomly draw $K=5$ representative clients for the federated experiments.

\subsection{Data Preprocessing}
\label{sec:preprocessing}
Because both datasets record discrete charging sessions rather than continuous time series, we first convert each session log into an hourly load profile. Each session's energy is distributed across the hours it spans in proportion to the minutes falling in each hour, assuming approximately constant charging power within a session \cite{ref_29}, after which hourly energies are summed over overlapping sessions and idle hours are set to zero. This conserves total session energy and avoids the artificial spikes of start-time aggregation. Palo Alto retains each station's full $14{,}462$-hour history, whereas EV~WATTS is aligned to a common $8{,}760$-hour ($365$-day) window. Each client's hourly load is then decomposed by CEEMDAN ($I=100$ trials, Eqs.~\eqref{eq:ceemdan_recon}--\eqref{eq:ceemdan_imf}) and zero-padded to a uniform $M=12$ IMF channels across the federation, which has no effect on training since zero channels contribute zero gradient. Finally, supervised sequences are formed with a $24$-hour lookback window ($\tau=24$, stride one) and split chronologically into $80\%$ training and $20\%$ test, with the per-IMF Min--Max scaler (Eq.~\eqref{eq:scale}) fitted on the training portion only to prevent temporal leakage. The resulting tensors have shape $(14{,}462, 24, 12)$ for Palo Alto and $(8{,}736, 24, 12)$ for EV~WATTS clients.

\subsection{Baselines}
\label{sec:baselines}

We benchmark SecFD-FL against two complementary families of
methods: (i) centralized deep-learning backbones trained on raw
and CEEMDAN-decomposed inputs, and (ii) representative
federated optimizers trained on the CEEMDAN-decomposed signal. All baselines share the same lookback window, train/test split, and optimization budget as SecFD-FL to ensure a controlled
comparison.

\subsubsection{Centralized backbones}
Five centralized backbones serves as baseline algorithms. LSTM stacks long short-term memory layers whose gated cell state retains long-range dependencies without the vanishing-gradient issue of vanilla recurrent neural networks \cite{ref_5}. GRU stacks Gated Recurrent Unit layers, replacing LSTM's separate cell state with a leaner update/reset gate design for faster convergence \cite{ref_30}. CNN applies stacked 1-D convolutional layers to capture local temporal patterns through shift-invariant filters rather than recurrence \cite{ref_5}. Seq2Seq with Attention couples an LSTM encoder--decoder with Bahdanau-style attention to mitigate the information bottleneck of a single context vector \cite{ref_32}. Transformer stacks self-attention blocks with positional encodings, replacing recurrence with parallel attention over the lookback window \cite{ref_5}.

\subsubsection{Federated optimizers}
We select six federated algorithms as baseline algorithms. FedAvg~\cite{FedAvg} aggregates client updates via a sample-size-weighted average and serves as the foundational baseline, though its convergence degrades under client drift. FedProx~\cite{Fedprox} adds a proximal term penalizing deviation from the global model to bound local drift under non-IID data. FedAdam~\cite{FedAdam} treats aggregated client updates as pseudo-gradients and applies server-side Adam-style momentum for more stable convergence. Per-FedAvg~\cite{Per_fed_avg} formulates federation as a MAML-inspired meta-learning problem, producing a global initialization that is rapidly personalized per client. SCAFFOLD~\cite{scaffold} introduces client and server control variates that correct local gradient drift, accelerating convergence under heterogeneity. FedDyn~\cite{feddyn} applies dynamic regularization via a per-client dual variable and a server-side error-feedback state to ensure convergence to the true global optimum.

\subsection{Evaluation Metrics}
\label{sec:metrics}

We evaluate SecFD-FL along three orthogonal axes: forecasting
utility, federated fairness, and privacy.

\subsubsection{Utility} For each client $k$ (an EV charging station or an individual EV), predicted load
$\hat{P}_k$ is compared to ground truth $P_k$ via Mean
Absolute Error (MAE), Root Mean Square Error (RMSE), and the coefficient of determination ($R^{2}$) as:
\begin{align}
\mathrm{MAE}_k&=\tfrac{1}{N}\!\sum_{i}|P_{k,i}{-}\hat{P}_{k,i}|,\\
\mathrm{RMSE}_k&=\sqrt{\tfrac{1}{N}\!\sum_{i}(P_{k,i}{-}\hat{P}_{k,i})^{2}},\\
R^{2}_{k}&=1-\tfrac{\sum_i(P_{k,i}-\hat{P}_{k,i})^2}{\sum_i(P_{k,i}-\bar{P}_k)^2}.
\end{align}
We report the federation-level mean and standard deviation of
each metric.

\subsubsection{Fairness} Performance equity across clients is
captured by Jain's index over per-client RMSE:
\begin{equation}
J=\frac{\bigl(\sum_{k=1}^{K}\mathrm{RMSE}_k\bigr)^{2}}
       {K\sum_{k=1}^{K}\mathrm{RMSE}_k^{2}},\quad J\in(0,1],
\label{eq:jain}
\end{equation}
where $J{=}1$ indicates perfectly equal client performance.

\subsubsection{Privacy} We instantiate the DLG attack of
Eq.~\eqref{eq:dlg} on both raw and SecAgg-masked updates
(300 attack iterations, lr $=0.1$). We report the
gradient-matching loss, the reconstruction MSE
$\mathrm{MSE}_\text{rec}=\!\|\mathbf{x}^{*}-\mathbf{x}\|_2^2/N$,
and the Pearson correlation
$\rho(\mathbf{x}^{*},\mathbf{x})$ between the recovered and
true samples; lower $|\rho|$ implies stronger protection.

\subsubsection{Communication overhead} Per-round upload is
$|\mathbf{w}|\!\cdot\!4$~bytes (float-32) per client, total
round cost is $2K|\mathbf{w}|\!\cdot\!4$~bytes
(upload + broadcast). The SecAgg mask overhead is reported
both in bytes (key-exchange only) and in milliseconds of
client-side compute.These quantities are evaluated in Section~\ref{sec:rq4}.

\subsection{Hyperparameter Settings}
\label{sec:hyper}

The hyperparameters used in all experiments are listed in Table~\ref{tab:hyper}. They were selected via a small grid search on a held-out validation split of one client and then fixed for both datasets. The frequency boundary $b{=}5$ was chosen so the residual stream ($\mathrm{IMF}_1$--$\mathrm{IMF}_6$) carries the high-frequency charging spikes while the trend stream ($\mathrm{IMF}_7$--$\mathrm{IMF}_{12})$ carries diurnal/seasonal regularities. We evaluate the frequency boundary at $b\in\{3,5,6,7\}$ on both datasets under identical training settings. On EV Watt, $R^2$ rises from $0.849$ at $b=3$ to $0.926$ at $b=5$, then falls slightly to $0.917$ and $0.910$ at $b=6$ and $b=7$. On Palo Alto, $R^2$ follows the same pattern ($0.782 \rightarrow 0.820 \rightarrow 0.799 \rightarrow 0.817$), confirming $b=5$ as the best setting on both datasets. The mask scale $\sigma_M{=}10^{3}$ guarantees an SNR $\!\approx\!10^{-6}$ from Eq.~\eqref{eq:snr}.

\begin{table}[!t]
\caption{Hyperparameters of the Proposed SecFD-FL Framework\label{tab:hyper}}
\centering
\renewcommand{\arraystretch}{1.2}
\begin{tabular}{|c|l|c|}
\hline
\textbf{Symbol} & \textbf{Description} & \textbf{Value}\\
\hline
$K$            & Number of clients                & 5 \\ \hline
$\tau$         & Lookback window (h)              & 24 \\ \hline
$M$            & IMF channels (zero-padded)       & 12 \\ \hline
$b$            & Frequency boundary index          & 5 \\ \hline
$T$            & Communication rounds              & 50 \\ \hline
$E$            & Local epochs / round              & 15 \\ \hline
$B$            & Mini-batch size                  & 64 \\ \hline
$\eta$         & Learning rate (Adam)              & $1{\times}10^{-3}$ \\ \hline
--             & Dropout                          & 0.20 \\ \hline
--             & GRU hidden units (layer 1 / 2)   & 64 / 32 \\ \hline
$\gamma$       & Local-retention ratio (resid.)   & 0.30 \\ \hline
$\sigma_M$     & Mask scale                       & $10^{3}$ \\ \hline
--             & Train/test split                  & 0.80 / 0.20 \\ \hline
\end{tabular}
\end{table}


\section{Results and Discussion}
\label{sec:results}

In this section, we present the performance results and comparison of the proposed SecFD-FL scheme with the aforementioned baseline algorithms in terms of (a) the effectiveness of CEEMDAN-driven frequency decoupling on forecasting accuracy; (b) the aggregation performance; (c) the defense performance against DLG; and (d) the communication and compute overhead.

\subsection{Effectiveness of Frequency Decoupling}
\label{sec:rq1}

Table~\ref{tab:cent} compares five centralized models trained on two types of inputs: (i) the raw hourly load (raw case) and (ii) the CEEMDAN-decomposed load (C-D case). First, it is observed that CEEMDAN-decomposed inputs consistently improve the performance of recurrent models (LSTM, GRU, Seq2Seq) and CNN on both datasets. For example, on the Palo Alto data, GRU reduces RMSE from $1.44$ to $0.80$~kWh ($-44.4\%$) and raises $R^2$ from $0.33$ to $0.79$. On EV Watt, GRU improves RMSE from $5.69$ to $3.55$~kWh ($-37.6\%$) and $R^2$ from $0.59$ to $0.79$. LSTM and Seq2Seq show comparable gains on both datasets. This pattern confirms that separating fast transient spikes from slow seasonal trends helps the model learn each component more effectively.

Second, among all models, GRU with CEEMDAN input achieves the best overall results. For the Palo Alto dataset, GRU records the lowest RMSE ($0.80$~kWh) and highest $R^2$ ($0.791$). On EV Watt, GRU again leads with RMSE of $3.55$~kWh and $R^2$ of $0.793$. This motivates our choice of GRU as the base
model in SecFD-FL. Recurrent models (LSTM, GRU, Seq2Seq) outperform CNN and Transformer because they are designed to capture sequential dependencies across time steps, which is important for EV charging patterns that depend on recent history. On the other hand, transformer performs poorly on both datasets, with negative $R^2$ values in most cases. This is expected because attention-based models lack a built-in preference for local temporal patterns and tend to overfit on short, noisy univariate signals \cite{new_cite_1}. CNN shows moderate improvement with CEEMDAN but still lags behind recurrent models due to its limited ability to model long-range dependencies.

\begin{table}[!t]
\caption{Centralized Backbones with Raw vs. CEEMDAN Inputs (Per-Station Mean over 5 Clients)\label{tab:cent}}
\centering
\renewcommand{\arraystretch}{1.2}
\setlength{\tabcolsep}{3pt}
\footnotesize
\begin{tabular}{|l|l|c|c|c|c|c|c|}
\hline
\textbf{Model} & \textbf{Input} & \multicolumn{3}{c|}{\textbf{EV Watt (user)}} & \multicolumn{3}{c|}{\textbf{Palo Alto (station)}} \\ \cline{3-8}
& & MAE & RMSE & $R^{2}$ & MAE & RMSE & $R^{2}$ \\ \hline
LSTM        & Raw  & 2.14 & 5.79 & 0.564 & 1.04 & 1.47 & 0.299 \\ \cline{2-8}
            & C-D  & 2.55 & 3.93 & 0.617 & 0.60 & 0.85 & 0.768 \\ \hline
GRU         & Raw  & 1.99 & 5.69 & 0.586 & 0.99 & 1.44 & 0.333 \\ \cline{2-8}
            & C-D  & 2.32 & 3.55 & 0.793 & 0.54 & 0.80 & 0.791 \\ \hline
Seq2Seq     & Raw  & 2.05 & 5.81 & 0.573 & 1.00 & 1.45 & 0.326 \\ \cline{2-8}
            & C-D  & 2.27 & 3.59 & 0.769 & 0.61 & 0.85 & 0.767 \\ \hline
Transformer & Raw  & 6.69 & 9.85 & $-$2.03 & 1.45 & 1.86 & $-$0.114 \\ \cline{2-8}
            & C-D  & 5.93 & 9.61 & $-$1.11 & 1.23 & 1.38 & 0.186 \\ \hline
CNN         & Raw  & 5.93 & 7.97 & $-$0.39 & 1.34 & 1.84 & $-$0.098 \\ \cline{2-8}
            & C-D  & 2.27 & 5.76 & 0.586 & 1.04 & 1.27 & 0.456 \\ \hline
\multicolumn{8}{l}{\footnotesize C-D: CEEMDAN-decomposed input; errors in kWh.} \\ 
\end{tabular}
\end{table}

\subsection{Aggregation Result Comparison}
\label{sec:rq2}

Table~\ref{tab:fl} compares the aforementioned FL benchmarks with the proposed SecFD-FL. We highlight three observation from the findings. 

\textbf{Heterogeneity-aware methods do not always help.} FedProx, SCAFFOLD, and FedDyn show poor performance on at least one client, producing negative $R^2$ values. For the EV Watt dataset, SCAFFOLD records an $R^2$ of $-0.529$ and RMSE of $9.82$~kWh, while FedDyn gives $R^2 = 0.248$ and RMSE of $6.63$~kWh. For the Palo Alto dataset, the same methods also lag behind FedAvg, with SCAFFOLD reaching only $R^2 = 0.181$. These methods apply corrections (proximal terms, control variates, or dynamic regularizers) that were designed for non-IID classification tasks. When applied to highly stochastic charging signals, these corrections can over-constrain the local updates and hurt convergence instead of helping it.

\textbf{Frequency decoupling improves federated learning.} By separating the model into residual and trend streams, SecFD-FL reaches a global RMSE of $2.28$~kWh and $R^2$ of $0.927$ on EV Watt, and $0.689$~kWh RMSE with $R^2$ of $0.847$ on Palo Alto. On EV Watt, standard FedAvg attains a marginally lower RMSE (2.08 vs.\ 2.28~kWh) and higher $R^2$ ( 0.938 vs.\ 0.927) than SecFD-FL; However, FedAvg offers no protection against gradient-inversion attacks, making SecFD-FL the preferable model when both forecasting accuracy and client-data privacy are required. The improvement comes from two sources: stream-specific aggregation weights match each stream's statistical nature, and local retention in the residual branch preserves client-specific transient patterns that would otherwise be averaged out. SecFD-FL achieves a Jain's fairness index of $0.871$ on EV Watt and $0.991$ on Palo Alto, the highest among all tested federated methods. This means that SecFD-FL maintains consistent performance across all clients, rather than favoring some at the expense of others.

\begin{figure}[!t]
\centering
\includegraphics[width=3.5 in]{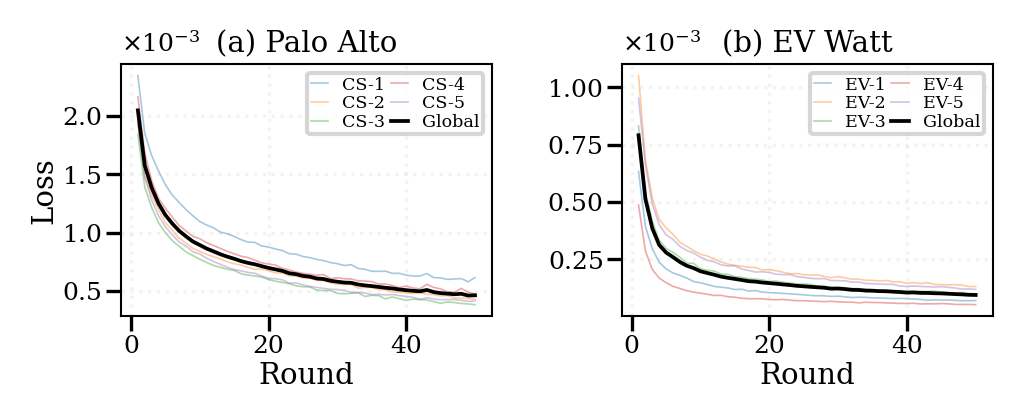}
\caption{SecFD-FL training convergence: per-client and globally averaged training loss over 50 rounds on (a) Palo Alto data (b) EV Watt data.}
\label{converge}
\end{figure}

\textbf{SecAgg does not reduce accuracy.}
Comparing FedAvg-FD (without privacy) and SecFD-FL (with SecAgg) in Table~\ref{tab:fl}, the two methods differ by less than $10^{-3}$ in every metric on both datasets. This confirms the result of Eq.~\eqref{eq:recover}: the zero-sum masks cancel exactly during aggregation, so SecFD-FL produces the same model as FedAvg-FD. The key difference is that SecFD-FL also protects client updates from the server, while FedAvg-FD does not. Fig.~\ref{converge} shows the training loss for all 5 clients across 50 rounds. The loss decreases smoothly and the global average stabilizes after approximately 25 rounds, showing that the masking protocol does not interfere with convergence.

\begin{table}[!t]
\caption{Federated Baselines and FD Variants (Global Metrics)\label{tab:fl}}
\centering
\renewcommand{\arraystretch}{1.2}
\setlength{\tabcolsep}{3pt}
\footnotesize
\begin{tabular}{|l|c|c|c|c|c|c|}
\hline
\textbf{Method} & \multicolumn{3}{c|}{\textbf{EV Watt}} & \multicolumn{3}{c|}{\textbf{Palo Alto}} \\ \cline{2-7}
& MAE & RMSE & $R^{2}$ & MAE & RMSE & $R^{2}$ \\ \hline
FedAvg              & 1.02 & 2.08 & 0.938 & 0.50 & 0.77 & 0.810 \\ \hline
FedProx             & 3.79 & 6.43 & 0.351 & 0.81 & 1.04 & 0.656 \\ \hline
FedAdam             & 1.88 & 2.87 & 0.861 & 0.56 & 0.83 & 0.781 \\ \hline
Per-FedAvg          & 1.72 & 2.72 & 0.875 & 0.53 & 0.77 & 0.808 \\ \hline
SCAFFOLD            & 6.66 & 9.82 & $-$0.529 & 1.29 & 1.60 & 0.181 \\ \hline
FedDyn              & 4.08 & 6.63 & 0.248 & 0.82 & 1.05 & 0.643 \\ \hline
FedAvg-FD           & 1.37 & 2.36 & 0.923 & 0.443 & 0.689 & 0.847 \\ \hline
\makecell[l]{\textbf{FedAvg-FD-SecAgg} \\ \textbf{or SecFD-FL (ours)}} & \textbf{1.22} & \textbf{2.28} & \textbf{0.927} & \textbf{0.443} & \textbf{0.689} & \textbf{0.847} \\ \hline
\end{tabular}
\end{table}

\subsection{Defense Performance against DLG}
\label{sec:rq3}

We evaluate the defence performance by running a 300-iteration DLG attack on the gradients of client~1. Two scenarios are tested:
(A) attack on raw gradients, as in standard FedAvg, and
(B) attack on masked gradients, as in SecFD-FL.
Table~\ref{tab:dlg} summarizes the results for both datasets.

Without SecAgg, the attacker achieves a best Pearson
correlation of $\rho = 0.689$ on EV Watt data and $\rho = 0.626$
on Palo Alto data. These values indicate that the attacker can
partially recover the original training data from the raw
gradients. The final gradient-matching loss remains low
($4.45$ on EV Watt, $2.18$ on Palo Alto), and the
reconstruction MSE stays small ($0.117$ and $0.078$),
meaning the attacker's reconstructed data stays close to the
true data throughout the attack.

With SecAgg enabled, the picture changes completely. The
best correlation drops to $\rho = 0.086$ on EV Watt and
$\rho = 0.040$ on Palo Alto, which are near zero and
statistically indistinguishable from random noise. The
gradient-matching loss increases by ten orders of magnitude
to $\sim\!4.55 \times 10^{10}$, showing that the attacker
cannot match the masked gradients at all. The reconstruction
MSE rises to $596.4$ and $601.4$, meaning the reconstructed
data is far from the original. Overall, SecAgg reduces the
reconstruction correlation by $8.1\times$ on EV Watt and
$15.6\times$ on Palo Alto.

\begin{figure}[!t]
\centering
\includegraphics[width=3.5 in]{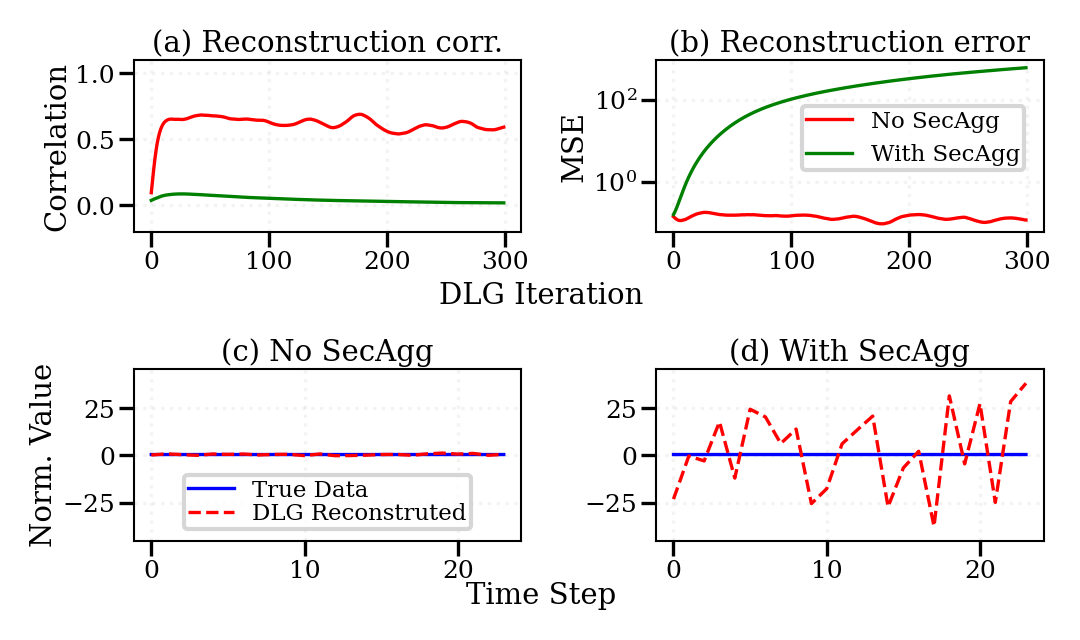}
\caption{DLG gradient-inversion attack on the EV Watt
dataset, Client~1. (a) Pearson correlation between
reconstructed and true samples over attack iterations.
(b) Reconstruction MSE over iterations.
(c) True vs.\ reconstructed IMF-1 sequence without SecAgg.
(d) True vs.\ reconstructed IMF-1 sequence with SecAgg.}
\label{dlg_analysis}
\end{figure}

Fig.~\ref{dlg_analysis} illustrates the DLG attack results
on the EV Watt dataset for client~1. Subplot~(a) shows the
Pearson correlation over attack iterations: for raw
gradients, the correlation rises quickly to around $0.6$ and
stays there, while for masked gradients, it remains near
zero. Subplot~(b) shows the reconstruction MSE: for raw
gradients, it stays low ($\sim\!0.1$); for masked gradients,
it grows steadily, confirming that the attacker moves further
from the true data at each step. Subplots~(c) and (d)
compare a sample IMF-1 sequence. Without SecAgg~(c), the
reconstructed signal closely follows the true data. With
SecAgg~(d), the reconstructed signal is random noise with
values ranging from $-30$ to $+40$, bearing no similarity to
the original. A similar analysis can be performed for the
Palo Alto dataset, with consistent conclusions as shown in
Table~\ref{tab:dlg}.

These results show that the zero-sum masking protocol
effectively prevents the server from recovering private
training data. By burying each client's update under
high-variance noise (SNR $\approx 2\times10^{-6}$, see
Eq.~\eqref{eq:snr}), the attacker can only match the noise
pattern rather than the true gradient. This ensures that
client data remains private throughout the federated
training process.

\begin{table}[!t]
\caption{DLG Gradient-Inversion Attack Outcomes\label{tab:dlg}}
\centering
\renewcommand{\arraystretch}{1.2}
\setlength{\tabcolsep}{3pt}
\footnotesize
\begin{tabular}{|l|c|c|c|c|}
\hline
\textbf{Quantity} & \multicolumn{2}{c|}{\textbf{EV Watt}} & \multicolumn{2}{c|}{\textbf{Palo Alto}} \\ \cline{2-5}
 & Raw & Masked & Raw & Masked \\ \hline
Final grad-match loss      & $4.45$  & $4.55{\times}10^{10}$ & $2.18$  & $4.55{\times}10^{10}$ \\ \hline
Final reconstruction MSE   & 0.117 & 596.4 & 0.078 & 601.4 \\ \hline
Correlation coefficient    & 0.689 & 0.086 & 0.626 & 0.040 \\ \hline
Reduction factor           & \multicolumn{2}{c|}{$\mathbf{8.1\times}$} & \multicolumn{2}{c|}{$\mathbf{15.6\times}$} \\ \hline
\end{tabular}
\end{table}

\subsection{Communication and Compute Overhead}
\label{sec:rq4}

The SecFD-FL model contains 46{,}860 trainable parameters. Using 32-bit floating-point representation, each parameter requires 4 bytes, giving a total model size of $183.05$~KB. In each communication round, every client uploads its masked model weights ($183.05$~KB) to the server, and the server broadcasts the updated global model ($183.05$~KB) back to each client. With $K = 5$ clients, the total upload per round is $5 \times 183.05 = 915.25$~KB, and the total download (broadcast) is also $915.25$~KB. This gives a total communication cost of $1{,}830.5$~KB per round and $89.4$~MB over the full 50 rounds of training.

The SecAgg masking protocol adds no additional \emph{per-round} communication overhead. Each client generates its mask locally each round using a pseudorandom number generator. It is seeded with the pairwise secrets established once during initialization. Since masks are computed on the client side, they are never transmitted over the network. The only setup cost is a one-time pairwise key exchange, which requires each pair of clients to exchange a 64-byte shared secret. For $K = 5$ clients, there are $K(K{-}1)/2 = 10$ pairs, giving a total key-exchange cost of $10 \times 64 = 640$~bytes ($0.625$~KB). This is negligible compared to the model traffic.

In terms of compute, mask generation takes $10.4 \pm 0.7$~ms per client per round, measured on a standard CPU. A single round of local training (15 epochs on $\sim$7000 samples) takes several seconds. Therefore, the mask generation adds less than $0.1\%$ to the total training time per round. Over 50 rounds, the total mask-generation time is approximately $0.52$~seconds. 

\subsection{Per-Station Forecast Quality}
\label{sec:rq5}

\begin{figure}[!t]
\centering
\includegraphics[width=3 in]{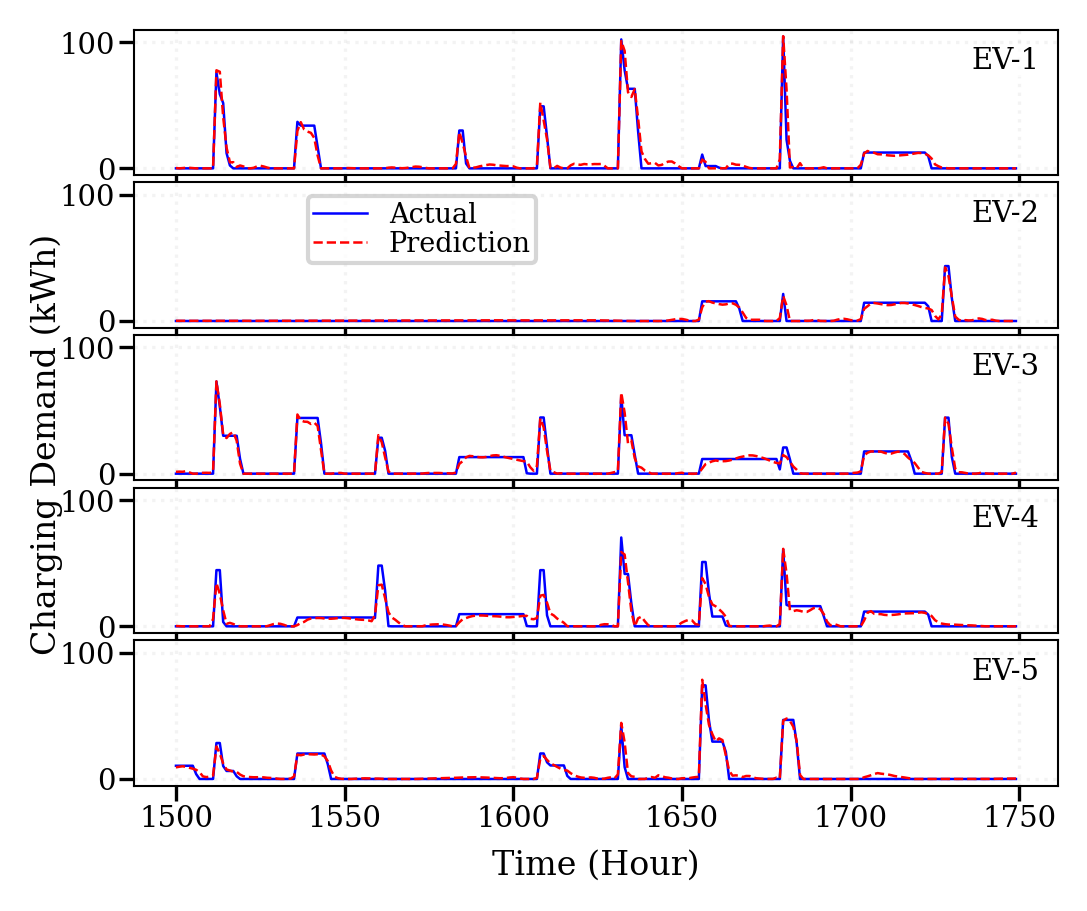}
\caption{True vs.\ predicted hourly energy demand for all
five clients on the EV Watt test set.}
\label{tru_pred}
\end{figure}

Fig.~\ref{tru_pred} shows the actual and predicted hourly
energy demand for each of the five EV clients on the EV Watt
dataset. The model accurately tracks both the long periods of
low or zero demand and the sudden high-energy charging events.
For example, client~1 has several peaks exceeding $100$~kWh,
and the predicted curve closely follows these spikes in both
timing and magnitude. A similar plot can be generated for the
Palo Alto dataset, where comparable tracking quality is
observed. Across all five clients, the global $R^2$ is $0.9265$.
Per-client $R^2$ values range from $0.8598$ (client~4) to
$0.9523$ (client~5), with clients~1, 2, 3, and 5 all
exceeding $0.93$. This narrow range shows that SecFD-FL
performs well across all clients, not just a few. The high
Jain's fairness index ($0.871$) reported in
Sec.~\ref{sec:rq2} further confirms this. SecFD-FL avoids the
common problem in federated forecasting where the global
model works well for some clients but poorly for others.

\section{Conclusion}
\label{sec:conclusion}

This paper introduced SecFD-FL, a federated learning framework for charging load forecasting in IoEV that couples CEEMDAN-driven frequency decoupling with stream-specific GRU model and a zero-sum secure-aggregation layer. By decomposing each client's raw load into high-frequency transients and low-frequency trends, and processing the two bands through a dual-stream GRU with
stream-specific aggregation rules, SecFD-FL attains forecasting accuracy on par with or superior to six representative FL algorithms across both user-level and station-level real-world datasets. The
embedded zero-sum masking protocol provides cryptographic-grade resistance against gradient-inversion attacks, 
reducing DLG reconstruction correlation by up to 15.6$\times$ while incurring no additional per-round communication (beyond a one-time 640-byte key exchange) and only $\sim$10 ms of per-round client compute.
Future work will extend SecFD-FL to dynamic client populations with asynchronous participation, integrate differential privacy guarantees on top of the masking layer, and investigate graph-aware aggregation for spatially correlated charging networks.

\section{Acknowledgments}

We acknowledged the useage of Claude and ChatGPT solely for language editing and grammer checking.

\bibliographystyle{IEEEtran}
\bibliography{references}

\end{document}